# Snapshot 3D image projection using a diffractive decoder

Çağatay Işıl[1,2,3,†], Alexander Chen[1,†], Yuhang Li[1,2,3,†], F. Onuralp Ardic[1], Shiqi Chen[1,2,3], Che-Yung Shen[1,2,3], Aydogan Ozcan[1,2,3,*]

**Affiliations**

[1]Electrical and Computer Engineering Department, University of California, Los Angeles, CA, 90095, USA

[2]Bioengineering Department, University of California, Los Angeles, CA, 90095, USA

[3]California NanoSystems Institute (CNSI), University of California, Los Angeles, CA, 90095, USA

\* ozcan@ucla.edu      [†] Equal contribution

**Abstract**

3D image display is essential for next-generation volumetric imaging; however, dense depth multiplexing for 3D image projection remains challenging because diffraction-induced cross-talk rapidly increases as the axial image planes get closer. Here, we introduce a 3D display system comprising a digital encoder and a diffractive optical decoder, which simultaneously projects different images onto multiple target axial planes with high axial resolution. By leveraging multi-layer diffractive wavefront decoding and deep learning-based end-to-end optimization, the system achieves high-fidelity depth-resolved 3D image projection in a snapshot, enabling axial plane separations on the order of a wavelength. The digital encoder leverages a Fourier encoder network to capture multi-scale spatial and frequency-domain features from input images, integrates axial position encoding, and generates a unified phase representation that simultaneously encodes all images to be axially projected in a single snapshot through a jointly-optimized diffractive decoder. We characterized the impact of diffractive decoder depth, output diffraction efficiency, spatial light modulator resolution, and axial encoding density, revealing trade-offs that govern axial separation and 3D image projection quality. We further demonstrated the capability to display volumetric images containing 28 axial slices, as well as the ability to dynamically reconfigure the axial locations of the image planes, performed on demand. Finally, we experimentally validated the presented approach, demonstrating close agreement between the measured results and the target images. These results establish the diffractive 3D display system as a compact and scalable framework for depth-resolved snapshot 3D image projection, with potential applications in holographic displays, AR/VR interfaces, and volumetric optical computing.

**Introduction**

Three-dimensional (3D) display has emerged as a foundational technology for next-generation holography, immersive visualization, and volumetric interfaces in AR/VR systems[1–5]. By delivering accurate focal cues across depth, 3D displays can provide more natural focus accommodation and alleviate vergence–accommodation conflict compared with conventional 2D screens, thereby improving depth perception and visual comfort. To achieve high-fidelity 3D volumetric display, a diverse range of paradigms has been explored, spanning volumetric displays based on microbubble voxels[6], acoustic trapping[7], or photophoretic optical trapping[8], light-field architectures such as lenticular lenslets and nanophotonic arrays[9], as well as holographic displays that directly encode wavefronts to synthesize objects in free space[10]. Among these, holographic displays have been widely investigated due to their inherent ability to display high-resolution 3D images by providing phase and amplitude control of the optical field[11–14]. However, achieving high axial resolution across multiple closely spaced depth planes remains a challenging task. As the axial image plane spacing decreases, diffraction-induced inter-plane coupling causes severe cross-talk, degrading depth selectivity and display fidelity[15]. These limitations stem from the insufficient degrees of freedom (DoF) available in conventional single-plane modulators, which cannot simultaneously satisfy the complex field requirements of multiple axial depths[10]. Consequently, practical holographic displays are bound by fundamental trade-offs between axial resolution, signal-to-noise ratio, and the space-bandwidth product of the system.

A wide range of computational holography algorithms has been developed to synthesize volumetric holograms with improved accuracy and color performance, forming an important algorithmic foundation of modern digital holographic displays[11,16–19]. More recently, learning-based methods have been incorporated into holographic and 3D display pipelines to enable data-driven wavefront generation, phase optimization, and end-to-end co-design, often improving display quality and computational efficiency[20–23]. While computational holography algorithms have significantly improved volumetric wavefront generation, their performance remains strictly governed by the hardware-limited DoF of spatial light modulators (SLMs). Recent optical augmentation strategies have introduced additional diffractive transformations to expand the effective DoF of the optical system, including active control of volume speckle fields and learned diffractive optics that extend system capability beyond an SLM-only pipeline[24–26].

Here, we present a snapshot 3D display system that integrates a digital encoder jointly optimized with a passive diffractive decoder composed of trainable phase layers. Within this framework, a learned diffractive decoder is co-designed with digital hologram synthesis through an encoder neural network to improve depth-dependent field shaping for snapshot 3D image projection over a desired volume. The digital encoder utilizes a Fourier-based network to extract multi-scale spatial and frequency features from input images, incorporates axial position information, and produces a single-phase representation that simultaneously encodes all input images for one-shot axial projection using a jointly optimized diffractive decoder. This end-to-end optimization, utilizing a diffractive decoder architecture, enables multi-plane snapshot 3D image projection, covering never-seen objects, with wavelength-scale axial separation between image planes while significantly suppressing inter-plane leakage/cross-talk compared to free-space-based image projection systems without a decoder. Stated differently, by embedding learned phase transformations into passive diffractive layers that serve as an optical decoder, the system physically performs depth-dependent field shaping during light propagation. This snapshot image projection architecture enables the optical routing of input images to specific depths while intrinsically suppressing inter-plane cross-talk, thereby moving beyond the inherent limitations of standard holographic image projection architectures.

We initially demonstrated this multi-plane snapshot image projection capability using four image planes with fine axial separation and further established the system's scalability through simulations of volumetric targets, achieving high-fidelity simultaneous image projection across 28 depth slices axially separated by one wavelength. Throughout our analysis, key performance factors, including the depth of the diffractive decoder, output diffraction efficiency, and encoded phase pixel count, were systematically evaluated to provide practical design guidelines for snapshot diffractive 3D displays. Furthermore, we investigated training strategies that enabled the dynamic adjustment of the axial position of the image projection plane across a continuous depth range rather than being restricted to fixed image planes at the output volume. Finally, we experimentally validated our snapshot image projection approach using a two-plane optical prototype operating at the visible part of the spectrum, where the measured intensity patterns closely matched our numerical simulations and target projections. Together, these results established a compact and scalable platform for high-axial-resolution snapshot 3D image display, advancing volumetric visualization and opening new opportunities for future holographic display technologies.

## Results

Our hybrid diffractive 3D display system consists of a digital encoder and an optical diffractive decoder that are jointly optimized. As shown in **Fig. 1A**, a 3D target volume was first decomposed into a stack of $M$ axial slices, augmented with a 2-channel positional grid, and processed by the digital encoder, which converted the volumetric input into a phase-only representation displayed on an SLM with $N_{encode} \times N_{encode}$ pixels. This encoded light field then propagated through a sequence of spatially-optimized diffractive surfaces that formed the physical decoder, shaping the output wavefront to display the desired image at its designated output volume. The digital encoder architecture is detailed in **Fig. 1B**, where a parallel Fourier encoder network[27–29] extracts multi-scale spatial and frequency-domain features from each plane, incorporates axial position encoding, and outputs an encoded phase representation that simultaneously represents all the desired images to be axially projected in a snapshot.

We first numerically demonstrate the 3D image projection capability of the presented diffractive display system. A digital encoder and a three-layer diffractive decoder were jointly optimized end-to-end to project test input images (never seen before) onto four discrete axial planes, as shown in the first row of **Fig. 2A**. The phase-modulating SLM plane contained $N_{encode} \times N_{encode}$ pixels, with $N_{encode} = 32$, while the output planes were also sampled with $32 \times 32$ pixels, while each diffractive layer comprises $256 \times 256$ diffractive features, with the optimized (and static) phase profiles shown in **Fig. 1C**. Unless otherwise specified, the lateral pixel pitch throughout this paper was set to $d = 0.6\lambda$. The axial spacing from the SLM to the first diffractive layer, between successive diffractive layers, and from the last diffractive layer to the first output plane was set to $30\lambda$. The four output image planes were uniformly separated by an axial distance of $\Delta z = 3\lambda$. The display results exhibit clear image projection at different axial image planes: each randomly selected handwritten digit (forming the input test images, which have never been seen before) is accurately displayed at its assigned axial distance while being effectively suppressed at the remaining image planes, confirming accurate image projection with minimal cross-talk across different image depths.

To investigate the impact of the number of diffractive layers used, we further trained two additional architectures for comparison: a single-layer diffractive decoder and a free-space-based baseline without a diffractive decoder. For each configuration, the digital encoder and the diffractive layers (where applicable) were jointly optimized[30–32]. Overall, the three-layer diffractive design achieved the highest image display fidelity, while the single-layer configuration produced distinguishable images with noticeable degradation. In contrast, the free-space baseline without a decoder failed to form well-resolved multi-plane image projections. We further quantitatively assessed the image projection fidelity of these architectures (for 10,000 test images never seen before) using the peak signal-to-noise ratio (PSNR) and the Pearson correlation coefficient (PCC), as summarized in **Fig. 2B** (see **Methods** for details). Averaged across the four output planes, the three-layer decoder achieved 22.5 dB PSNR and 0.92 PCC, outperforming the single-layer decoder (22.0 dB, 0.91) and the free-space baseline (19.8 dB, 0.79). These results confirm that increased architectural depth of the diffractive decoder substantially improves the 3D image projection fidelity and enables precise axial separation of closely spaced output planes.

To further evaluate the spatial-frequency handling capability of the diffractive 3D display system, we performed blind testing using a set of tightly spaced line patterns with a period of $1.2\lambda$, as shown in **Fig. 3**. These grating patterns, including vertical and horizontal line-pairs as well as their combinations across axial planes, were never used during training and served as an external test dataset. The three-layer diffractive decoder successfully displayed the paired gratings at their designated axial planes, while strongly suppressing cross-talk and their leakage at other depths. In comparison, the single-layer decoder system yielded partial reconstructions with noticeable blur and axial cross-talk, whereas the free-space baseline failed to resolve the grating pairs and had major cross-talk across different image planes.

We then evaluated the performance trade-off between the output diffraction efficiency and the display fidelity of the system by introducing an energy efficiency constraint to the training loss function, while maintaining the same four-channel input configuration as in **Fig. 2**. These 3D display systems were optimized using a composite reconstruction error alongside a diffraction efficiency penalty term (see the **Methods** for details). As shown in **Fig. 4A,** the PSNR and PCC values decreased as the output diffraction efficiency increased, revealing a clear efficiency–fidelity trade-off: higher output diffraction efficiency improved overall image brightness but also introduced stronger interference artifacts and axial cross-talk, which reduced display fidelity. Representative image projections at three output efficiency regimes: low ($\sim$0.0005–0.01%), moderate ($\sim$10–11%), and high ($\sim$30–35%), are shown in **Fig. 4B**. At low diffraction efficiencies, the display results were structurally clean, whereas increasing the efficiency introduced visible speckle noise. We further compared free-space propagation, single-layer, and three-layer diffractive decoder architectures under matched efficiency regimes. Across all regimes, the three-layer decoder configuration consistently delivered the best 3D image display quality, followed by the single-layer decoder design, while the free-space propagation without the diffractive decoder yielded the lowest performance. Notably, even in the lowest diffraction efficiency regime, the free-space propagation-based decoder exhibited inferior performance, with strong cross-talk observed between axial planes, underscoring the fundamental advantages of increased depth of the diffractive decoder.

Building on the preceding analysis, which focused on multi-plane image projection for a few discrete axial planes, we extended our method to a continuous volumetric image projection scenario involving a substantially larger number of axial slices. To accommodate the increased axial density, we trained a new model specifically optimized for this higher-resolution volumetric image projection task (see the **Methods** for details). **Figure 5A** reports the diffractive 3D display capability of our system using a 28-slice airplane model (a test object never seen before), which imposed a higher axial workload on the system. The 3D volume was decomposed into 28 axial slices with a uniform separation of $\Delta z = 1\lambda$, and these axial slices were encoded to a single-phase pattern for simultaneous projection in a snapshot. During inference, this single encoded hologram reconstructed each slice at its assigned axial depth through the optimized diffractive decoder, enabling volumetric projection across an extended axial range. As illustrated in **Fig. 5B**, the system successfully reproduced the main structural components of the 'airplane' across all 28 axial planes. Slightly reduced fidelity was observed near the central slices, consistent with increased geometric complexity, while outer slices exhibited visually clearer results. Quantitative metrics in **Fig. 5C** further confirmed this trend: both PSNR and PCC values showed a dip near the mid-slices, followed by improvement toward the volume boundaries. Together, these results demonstrate that the proposed snapshot 3D diffractive display can scale from a few-plane image projection to large volumetric workloads, supporting multi-slice 3D rendering in a snapshot using a single input phase encoding.

To further dissect the factors affecting diffractive 3D display quality, we examined the interplay between the number of encoder pixels and the presence of a learned diffractive decoder. Specifically, we compared two SLM configurations parameterized as $N_{encode} \times N_{enocde}$, with $N_{encode} \in \{32, 192\}$, under two propagation conditions: free-space-based propagation and a single-layer diffractive decoder. In these analyses, the pixel pitch was fixed to $d = 1\lambda$. As shown in **Fig. 6A**, increasing the number of encoder pixels $N_{SLM} = N_{encode}^2$ markedly improved the perceptual sharpness of the test images (never seen before) under free-space propagation; however, these reconstructions still exhibited pronounced inter-plane leakage due to limited modulation capability. Coupling the higher DoF phase pattern ($N_{encode} = 192$) with a single diffractive layer-based decoder (256×256 features) yields a substantial improvement in 3D image projection fidelity. As shown in **Fig. 6B**, the overall PSNR increases from 19.8 dB ($N_{encode} = 32$, free-space) to 21.4 dB ($N_{encode} = 192$, free-space) and reaches 23.2 dB with the diffractive decoder, while the PCC metrics improve from 0.79 to 0.88 and 0.92, respectively. These results indicate that higher-resolution phase modulation alone is insufficient for depth-multiplexed 3D image display; they also underscore the critical role of an optimized diffractive decoder, which is essential for suppressing axial cross-talk and decoupling adjacent axial planes from each other, thereby enabling a high-fidelity 3D display that higher-resolution phase modulation alone cannot achieve.

We further assessed the ability of the diffractive 3D display to dynamically tune the target projection plane. To this end, we varied the target axial offsets $\delta z^*$ used to condition the digital encoder, as shown in **Fig. 7A**. This strategy enables the encoder digital network to dynamically generate phase patterns for different desired axial locations. During training, we considered four axial sampling strategies over $[0, \Delta z]$: discrete sampling at 3, 5, or 8 uniformly spaced axial planes, and continuous sampling where $\delta z^* \sim U(0, \Delta z)$, where $\Delta z = 3\lambda$. The continuous axial sampling model was initialized from the 8-position model via transfer learning to accelerate its convergence. We blindly tested all models over a continuous sweep of axial shifts, $\delta z/\Delta z \in [0,1]$, which was used as an input to the encoder network to set the desired axial projection distance, as shown in **Fig. 7B**. Sparse axial encoding (3 or 5 axial positions) led to pronounced, oscillatory behavior in the output PSNR, with sharp degradations at intermediate depths that were not used during training. In contrast, denser discrete sampling and continuous conditioning of the digital encoder yielded smooth, consistent performance across the entire axial range, thereby improving the overall 3D image display fidelity. Representative results in **Fig. 7C** corroborate these trends: sparsely trained models exhibit blur and geometric distortions at intermediate axial offsets, whereas the continuously trained model maintains sharp, structurally stable projections throughout $[0, \Delta z]$. Collectively, these results indicate that denser axial sampling during training is crucial for empowering the system with robust axial position adjustment capabilities, while also improving overall image display performance.

Beyond these numerical analyses, we also experimentally validated the presented diffractive 3D display system. We jointly trained a digital encoder with a single-layer physical decoder to simultaneously project test images (never seen before) onto two axial image planes, as illustrated in **Fig. 8A**. In this configuration, two test images were first processed by the digital encoder to generate a phase-only pattern, which was then displayed on SLM1. As shown in the schematic in **Fig. 8B**, coherent illumination ($\lambda = 650$ nm) interacted with this phase modulation, propagated through the optical path, and was subsequently decoded by SLM2, implementing the learned physical decoder function. A photograph of the experimental setup is provided in **Fig. 8C**. **Figure 8D** visualizes the target images, numerical simulations, and experimentally captured results at the $1^{st}$ and $2^{nd}$ axial planes, separated by $\Delta z = 1$ cm. The experimental outputs closely match the corresponding simulations and target patterns across a diverse set of handwritten digits (never seen before), demonstrating that the trained hybrid digital–physical architecture was accurately realized in the actual optical system operating in the visible spectrum. These results confirm the robustness of the learned design and the physical decoder's ability to faithfully reproduce multi-plane image projections. Additional experimental results are also provided in **Supplementary Fig. S1**, further supporting the same conclusions.

## Discussion

In this work, we introduced a hybrid 3D diffractive display system that jointly optimizes digital hologram synthesis and multi-layer diffractive decoding to improve axial plane separability in depth-multiplexed 3D image projection in a snapshot. By training a digital encoder together with an optical diffractive decoder, the system learned to simultaneously route different test images to distinct axial planes and suppress their leakage across depth. Our results demonstrate that the multi-layer configuration consistently outperforms single-layer and free-space baselines, showing that diffractive depth is essential for accurate axial routing and suppression of inter-plane leakage, in the sense that additional controllable degrees of freedom during propagation can be used to route energy to designated axial planes that are densely packed. The framework also generalizes beyond the training distribution, successfully displaying, for example, unseen high-frequency 3D patterns. Additionally, we demonstrated the scalability of this approach to larger volumetric workloads by displaying a 28-slice volumetric scene with an axial separation of $\Delta z = 1\lambda$ between two successive image planes.

Our ablation studies also provide practical design guidance for hybrid diffractive 3D displays. We systematically investigated key factors that govern system performance, including diffraction efficiency, the number of encoder pixels used for phase display, and the ability to dynamically adjust the target projection planes. An explicit trade-off between the output diffraction efficiency and 3D image display fidelity was observed, where at higher diffraction efficiencies, we observed stronger artifacts and increased cross-talk. We also showed that while increasing the encoder pixel counts controlled by the encoder network improved image sharpness, it could not eliminate depth leakage on its own; incorporating a learned diffractive decoder (jointly optimized with the encoder network) provided significant gains beyond the high-resolution free-space case, indicating that the optical decoder contributed a complementary and essential depth-selective snapshot 3D image projection capacity. Regarding depth-tunable 3D image displays, depth conditioning of the encoder enabled the dynamic adjustment of target image planes; furthermore, dense axial sampling during training was found to be crucial for empowering the system with robust image depth adjustment capabilities and ensuring smooth performance across intermediate image depths. Taken together, our results underscored the importance of joint digital–optical optimization, where computational encoding and physical wavefront shaping operated in synergy to maximize the effective space–bandwidth product of the overall snapshot 3D image projection system. Finally, we experimentally validated the proposed framework, where the measured images closely matched the simulations and successfully reproduced the target images, confirming the feasibility of the hybrid diffractive system for snapshot 3D image display.

While the presented diffractive 3D display system demonstrates strong volumetric projection capabilities, several challenges remain. Realizing multi-layer diffractive decoders places an increasing demand on precise lateral and axial alignments and stability as the layer count increases; this motivates vaccination-based training strategies, where perturbations are intentionally introduced during training to improve tolerance to fabrication and alignment errors[33]. Extending the approach to broadband or color operation represents another direction, which would require joint optimization across wavelengths and careful consideration of material dispersion in the decoder layers[11,34]. Furthermore, scaling to higher spatial resolution or dynamic volumetric scenes would increase training and memory costs due to the requirement for larger propagation models and finer target grids, necessitating the adoption of efficient parallel computing techniques and distributed optimization strategies[35]. Consequently, continued advances in 3D fabrication of diffractive decoders, calibration procedures and *in situ* learning methods[36,37], as well as model-based training and higher-resolution modulators, would be essential for translating the presented framework into practical systems.

Looking forward, the presented diffractive snapshot 3D display framework opens opportunities for compact holographic displays, multi-depth AR/VR projection, volumetric microscopy, and real-time 3D visualization systems. Its ability to encode and route information across depth suggests potential extensions to multispectral operation and multi-perspective holography. We anticipate that this hybrid approach will contribute to the development of the next generation of high-fidelity 3D display technologies.

## Methods

### Digital encoder design

As illustrated in **Fig. 1B**, we employed a parallel Fourier encoder network as the computational front end to synthesize the snapshot wavefront needed for 3D image projection. The network took a target 3D volume, represented as an $M$-channel stack of 2D intensity slices, and output a phase-encoded single pattern that modulated the incident field to display the desired volumetric image at the specified output volume. To support depth-adaptive and spatially-aware phase encoding, we augmented the input with additional spatial context channels. First, a 2-channel normalized Cartesian coordinate grid $(x, y)$ was appended to the input slices, allowing the translation-invariant convolutional kernels to learn accurate wavefront shaping required for 3D image projection. Second, inside the network, the target axial shift $\delta z^*$ was spatially broadcast into a depth plane map and further concatenated with the input, conditioning the network on the desired axial projection depth. The digital encoder comprises convolutional layers coupled with multiple Fourier modules that operate in parallel. This multi-branch design enabled the network to capture complementary global and local image features. The Fourier modules operated in the frequency domain, learning global filters and modeling long-range dependencies. The real and imaginary components of the frequency representation were processed separately and then recombined into real-valued feature maps. Convolutional blocks with varying kernel sizes extract robust, multi-scale local features in the spatial domain. These global and local feature maps were then concatenated and passed through a fully connected layer to produce the final encoded phase-only pattern for the snapshot 3D image projection.

### Physical decoder design

The physical decoder consists of a sequence of passive and stationary diffractive layers positioned after the encoder spatial light modulator. We modeled the system using monochromatic illumination with wavelength $\lambda$. The optical forward model was expressed as a sequence of free-space propagations and phase modulations. Free-space propagation between two consecutive planes separated by an axial distance $z_d$ was modeled using the angular spectrum approach:

$$u(x, y, z + z_d) = \mathcal{F}^{-1}\{\mathcal{F}\{u(x, y, z)\} \cdot H(f_x, f_y; z_d)\} \tag{1}$$

where $u(x, y, z)$ denotes the complex optical field at plane $z$ along the optical axis, and $u(x, y, z + z_d)$ is the resulting propagated field after the distance $z_d$. $f_x$ and $f_y$ are the spatial frequencies along the $x$- and $y$- directions, respectively. $\mathcal{F}$ and $\mathcal{F}^{-1}$ represent the two-dimensional Fourier and inverse Fourier transforms, and $H(f_x, f_y; z_d)$ is the free-space transfer function:

$$H(f_x, f_y; z_d) = \begin{cases} \exp\left\{jkz_d\sqrt{1 - \left(\dfrac{2\pi f_x}{k}\right)^2 - \left(\dfrac{2\pi f_y}{k}\right)^2}\right\}, & f_x^2 + f_y^2 < \dfrac{1}{\lambda^2} \\ 0, & f_x^2 + f_y^2 \geq \dfrac{1}{\lambda^2} \end{cases} \tag{2}$$

where $j = \sqrt{-1}$ and $k = \frac{2\pi}{\lambda}$.

Each diffractive layer was modeled as a thin phase-only modulation element with negligible absorption. The transmission coefficient of the $k^{th}$ layer located at $z = z_k$ is:

$$t(x, y) = \exp\{j\phi(x, y, z_k)\} \tag{3}$$

where $\phi(x, y, z_k)$ is the trainable phase modulation value. These phase values constituted the learnable parameters of the 3D image projection diffractive decoder.

In the experimental implementations of the diffractive projection system, misalignments between the diffractive layers may degrade the projection quality. To enhance robustness, we introduced random misalignments during training. The lateral and axial displacements $D = (D_x, D_y, D_z)$, were introduced as independent, uniformly distributed random variables. These displacement errors $D_x$, $D_y$, and $D_z$ were modeled by uniformly distributed random variables,

$$D_x \sim U(-\Delta_x, \Delta_x), D_y \sim U(-\Delta_y, \Delta_y), D_z \sim U(-\Delta_z, \Delta_z) \tag{4}$$

where $\Delta_x$, $\Delta_y$, and $\Delta_z$ represent the maximum displacements along the $x$-, $y$-, and $z$- axes, respectively. In our designs, we set the misalignment bounds uniformly as $\Delta_{max} = \Delta_x = \Delta_y = \Delta_z = 0.5d$, where $d = 8$ μm is the pixel pitch for the experimental setup.

**Implementation details for numerical analysis**

Numerical simulations of the diffractive 3D image display system were performed under coherent monochromatic illumination at a wavelength of $\lambda = 0.75\ mm$. During the simulations, the input images were grayscale and resized to 32×32 pixels before being fed into the network, following the digital encoder and physical decoder described above. Each diffractive layer was composed of $256 \times 256$ trainable phase features ($153.6\lambda \times 153.6\lambda$), with a lateral feature size of $d = 0.6\lambda$. The central $32 \times 32$ region ($19.2\lambda \times 19.2\lambda$) of the sensor plane was used for loss function calculation. In Fig. 6, we fixed the pixel pitch to $d = 1\lambda$ to improve training stability for the larger SLM configuration, while keeping all the other settings the same.

The diffractive 3D display systems were trained using Python (v3.10.6) and JAX/Flax framework, utilizing the Optax library for optimization, with the Adam optimizer and a learning rate of $1 \times 10^{-4}$ for the digital encoder and $1 \times 10^{-3}$ for the diffractive decoder. The batch size was set to 64, and training was run for 2000 epochs. Diffractive layer phase values were initialized to zero. The training was performed on a GeForce GTX 4090 GPU (Nvidia Corp.), and each training run typically required ~24 hours.

**Implementation details for experimental results**

Experimental validation of the snapshot 3D image display system was performed using the optical setup illustrated in **Fig. 8C**. The system utilized a coherent monochromatic light source (Fianium) operating at a wavelength of $\lambda = 650$ nm. The experimental system consisted of a digital encoder plane and a single diffractive decoder layer separated by an axial distance of $z = 10$ cm. The encoded phase patterns had $512 \times 512$ pixels, and the diffractive decoder had $900 \times 900$ pixels. The optical path was folded using two beam splitters (BS1 and BS2) to accommodate a two-stage modulation process involving two reflective SLMs (i.e., SLM1 and SLM2). The digital encoder first processed the target input images to generate the holographic phase pattern. This encoder pattern was displayed on SLM1 (HOLOEYE PLUTO-2.1; pixel pitch, 8 μm; resolution, $1,920 \times 1,080$), which served as the digital-to-optical interface, modulating the incident planar wavefront to create the encoded light field. The modulated field from SLM1 was directed by BS2 to SLM2 (HOLOEYE LUNA; pixel pitch, 4.5 μm; resolution, $1,920 \times 1,200$), which was positioned at an axial distance of $z = 10$ cm from SLM1. SLM2 acted as the physical decoder; it displayed the learned phase profile of the diffractive layer, effectively functioning as the passive, static diffractive surface described in the forward model. After modulation by the physical decoder, the light field propagated to the output region. A CMOS image sensor (Basler ace acA1920-40gm; pixel pitch, 5.86 $\mu m$; resolution, $1,920 \times 1,200$) was used to capture the resulting volumetric intensity distribution. A central $600 \times 600$ pixel region of the sensor was recorded and used for analysis. To evaluate the multi-plane display capability, the camera was mounted on a translation stage to record intensity measurements at multiple axial planes (e.g., $1^{st}$ Plane and $2^{nd}$ Plane as shown in **Fig. 8A**) corresponding to the depths of the target objects.

**Training and test datasets**

We used the MNIST[38] handwritten digits dataset for training and testing of most of the diffractive 3D display systems reported in this work. Specifically, 50k handwritten images were used for training, and the remaining 10k were reserved for testing. Individual digit images were randomly selected and assigned to different axial planes, and random combinations of digits were used to construct synthetic multi-plane targets for both training and testing. This dataset enables systematic evaluation of axial image routing, inter-plane cross-talk suppression, and generalization performance under well-controlled and widely adopted benchmarks. For the numerical simulations involving higher axial complexity, specifically the 28-slice volumetric image projection task, we used the 3D ShapeNets dataset[39]. Voxelized three-dimensional objects were represented as stacks of axial slices and used as depth-resolved supervision for training and blind testing. This dataset enabled the assessment of volumetric display performance for the presented system across a larger number of axial planes.

**Supplementary Information:**

- Training loss function and performance evaluation metrics

- Supplementary Figure 1

# Figures

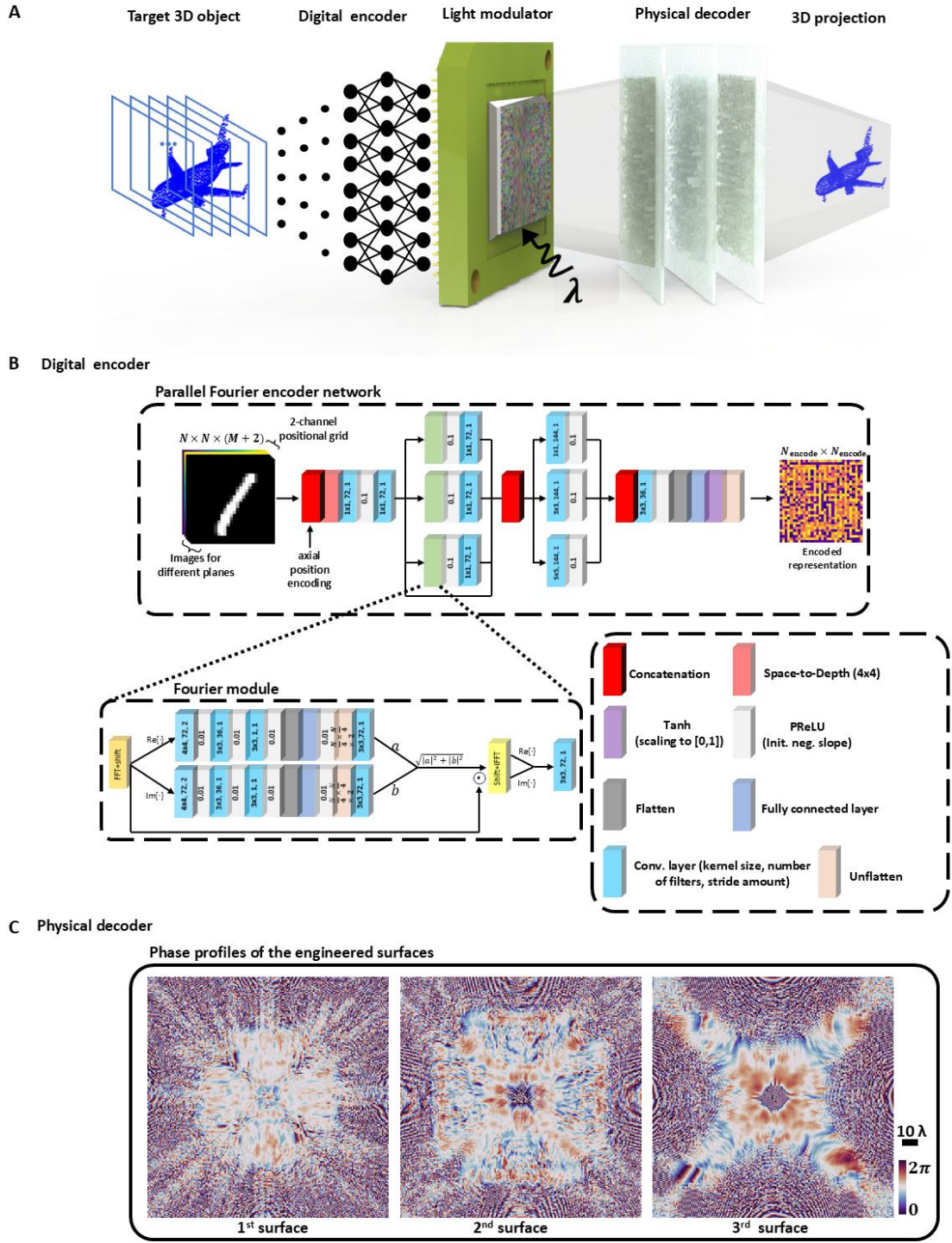

**Figure 1. Schematic of the diffractive snapshot 3D display architecture. A,** Conceptual illustration of the diffractive snapshot 3D display system, consisting of a digital encoder and a physical decoder. **B,** Structure of the digital encoder. **C,** Phase profiles of the optimized diffractive surfaces constituting the physical decoder.

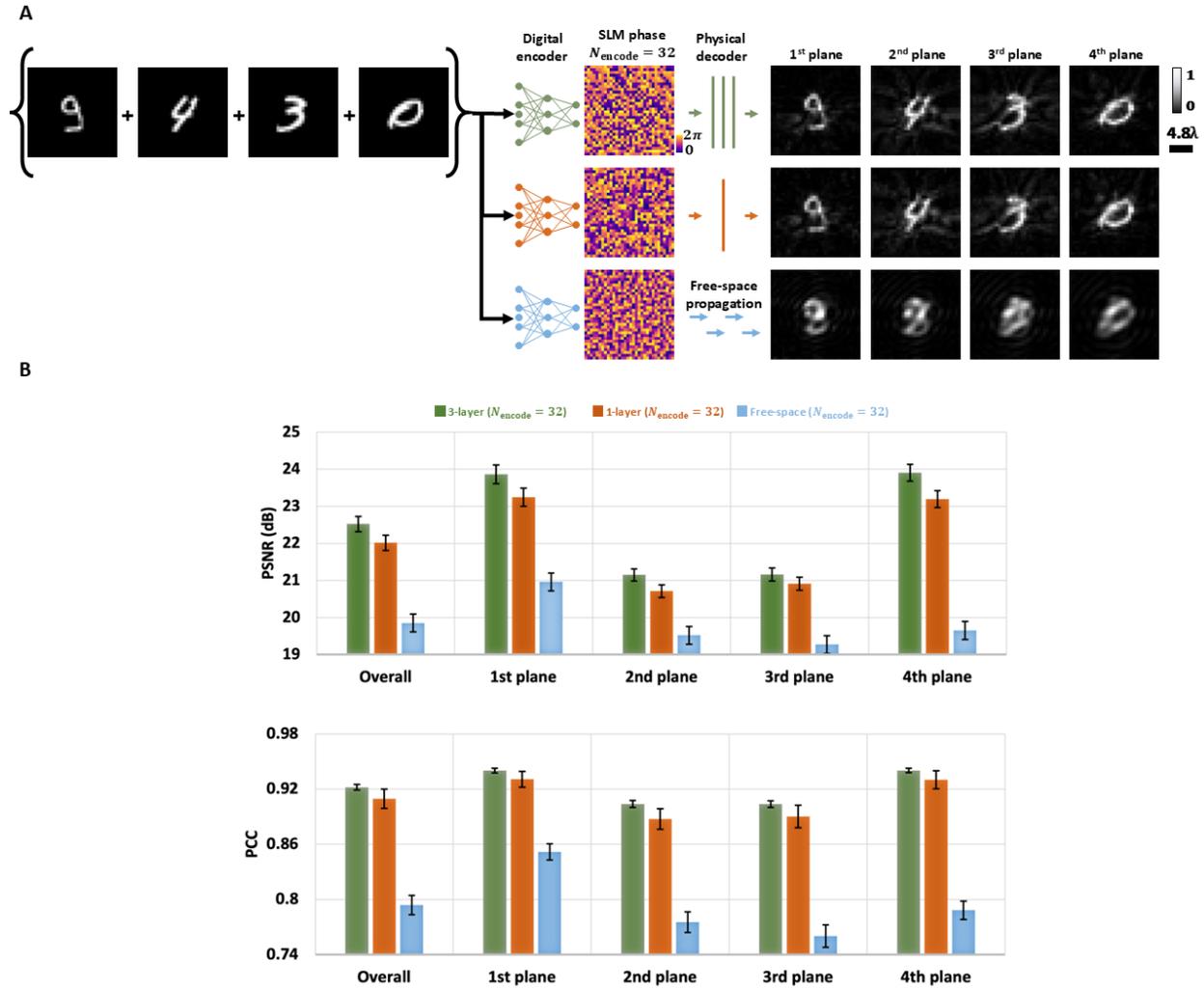

**Figure 2. Numerical analysis of the diffractive snapshot 3D display system. A**, Representative examples of snapshot multi-plane image projection, where different image content was displayed at four distinct axial planes (separated by $\Delta z = 3\lambda$), reported for a three-layer diffractive decoder, a single-layer diffractive decoder, and a free-space baseline used for comparison. **B**, Quantitative image display fidelity evaluated on the test image set (never seen before) using PSNR and PCC metrics for the three optical configurations. The multi-layer diffractive 3D display provides better image display fidelity at each projection plane.

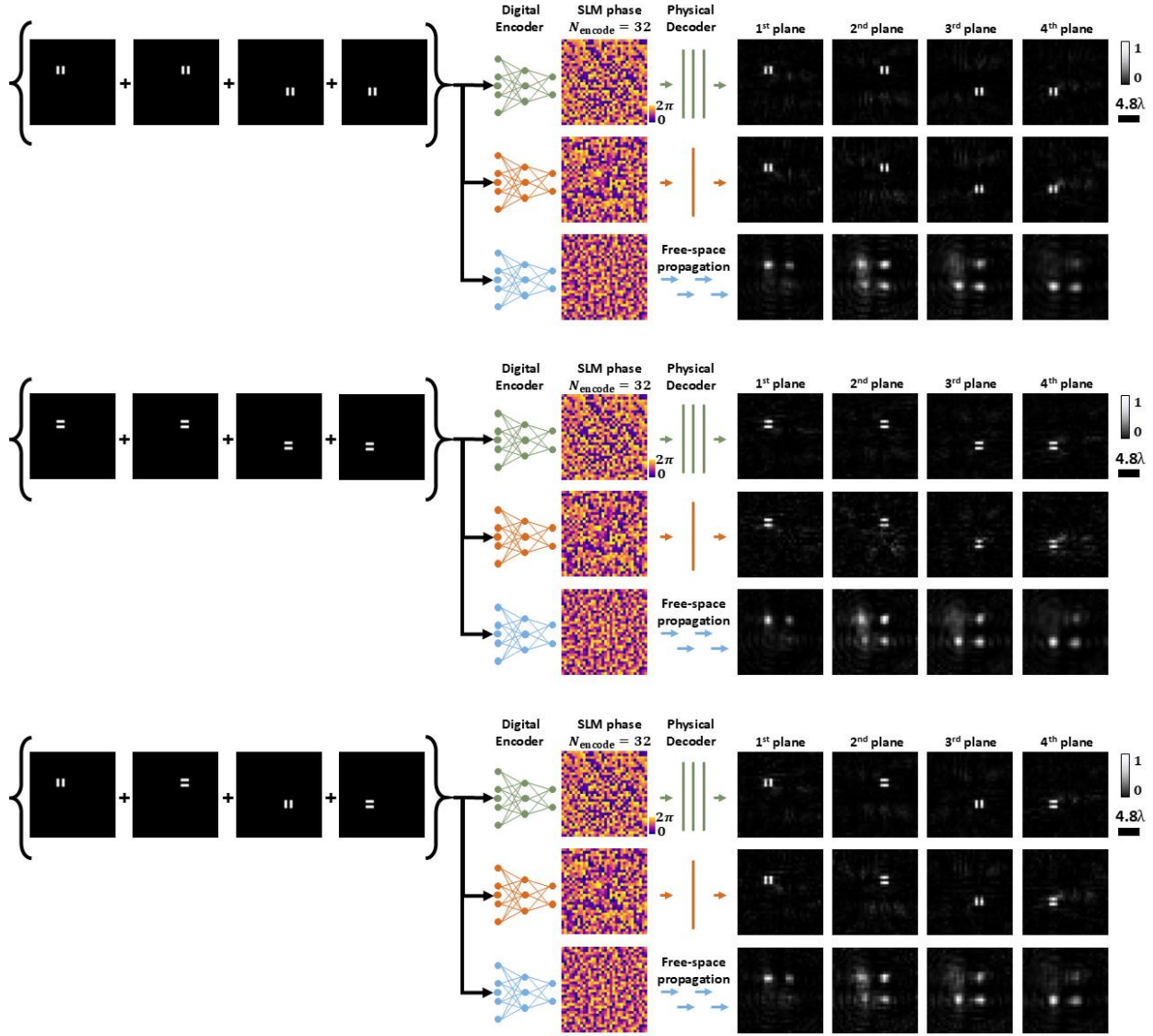

**Figure 3. Multi-plane snapshot projection of line patterns using the diffractive 3D display.** Blind testing using vertical, horizontal, and mixed bar patterns, each with a period of $1.2\lambda$. The mixed bar patterns correspond to configurations in which vertical and horizontal bars were distributed across different axial planes. These line patterns, including vertical, horizontal, and mixed structures, were never used during training and serve as an external test dataset. Three groups of images are shown, each consisting of four images corresponding to four target axial planes, for a three-layer diffractive decoder, a single-layer diffractive decoder, and the free-space baseline. The multi-layer diffractive 3D display accurately projected the line pairs at their designated axial planes, whereas the single-layer decoder configuration exhibited blur, and the free-space setup failed to display the line pairs, also exhibiting significant cross-talk among different planes. Each axial plane was separated from the neighboring planes by $\Delta z = 3\lambda$.

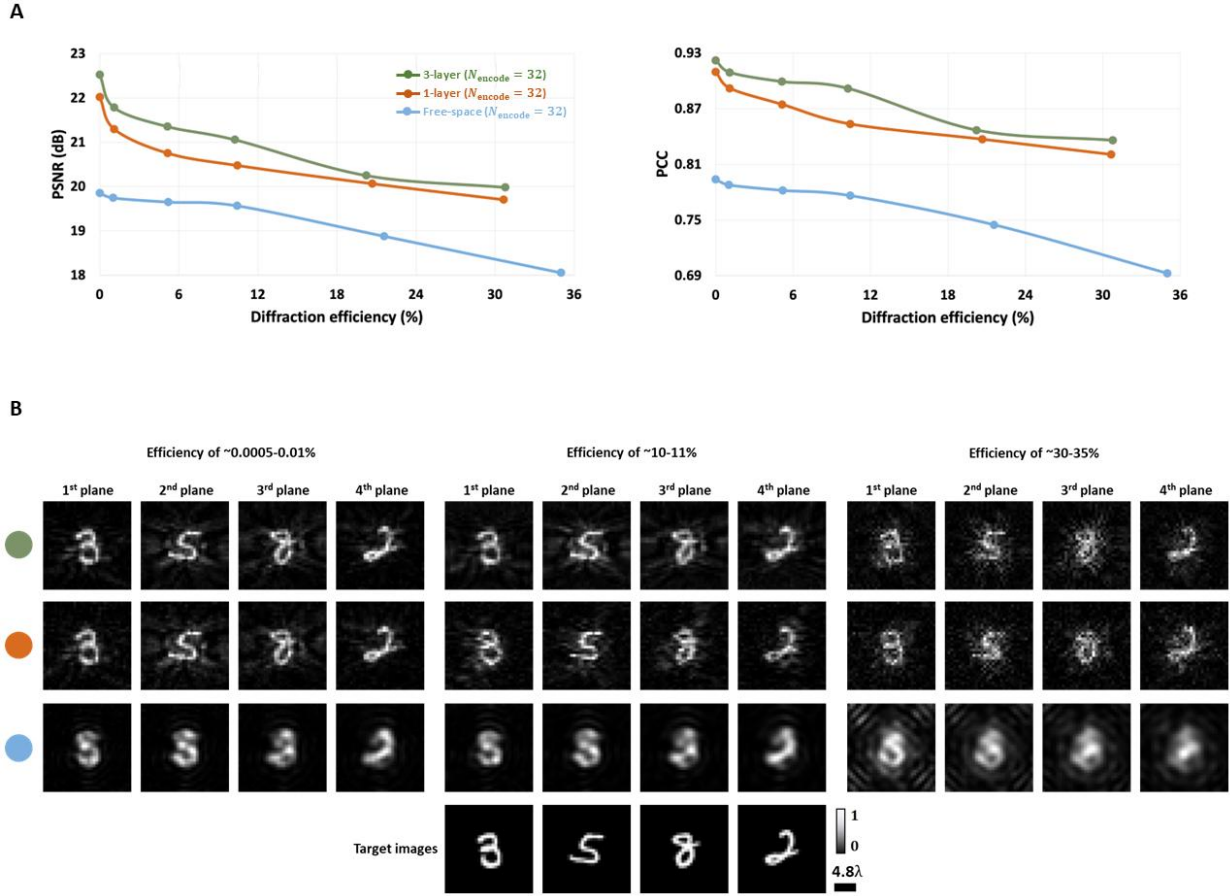

**Figure 4. Diffraction efficiency vs. performance trade-off analysis of the snapshot diffractive 3D display. A**, Output PSNR and PCC values as a function of the diffraction efficiency for three optical configurations: a three-layer diffractive decoder, a single-layer diffractive decoder, and the free-space baseline, showing an inherent efficiency–performance trade-off. **B**, Representative multi-plane projections at low, moderate, and high diffraction efficiency levels. Low-efficiency outputs were clean, moderate efficiency provided balanced contrast, while high diffraction efficiency introduced increased speckle and inter-plane cross-talk. The colored circles on the left correspond to the same three-layer diffractive decoder, single-layer diffractive decoder, and the free-space baseline reported in **A**. Axial separation between image planes was $\Delta z = 3\lambda$.

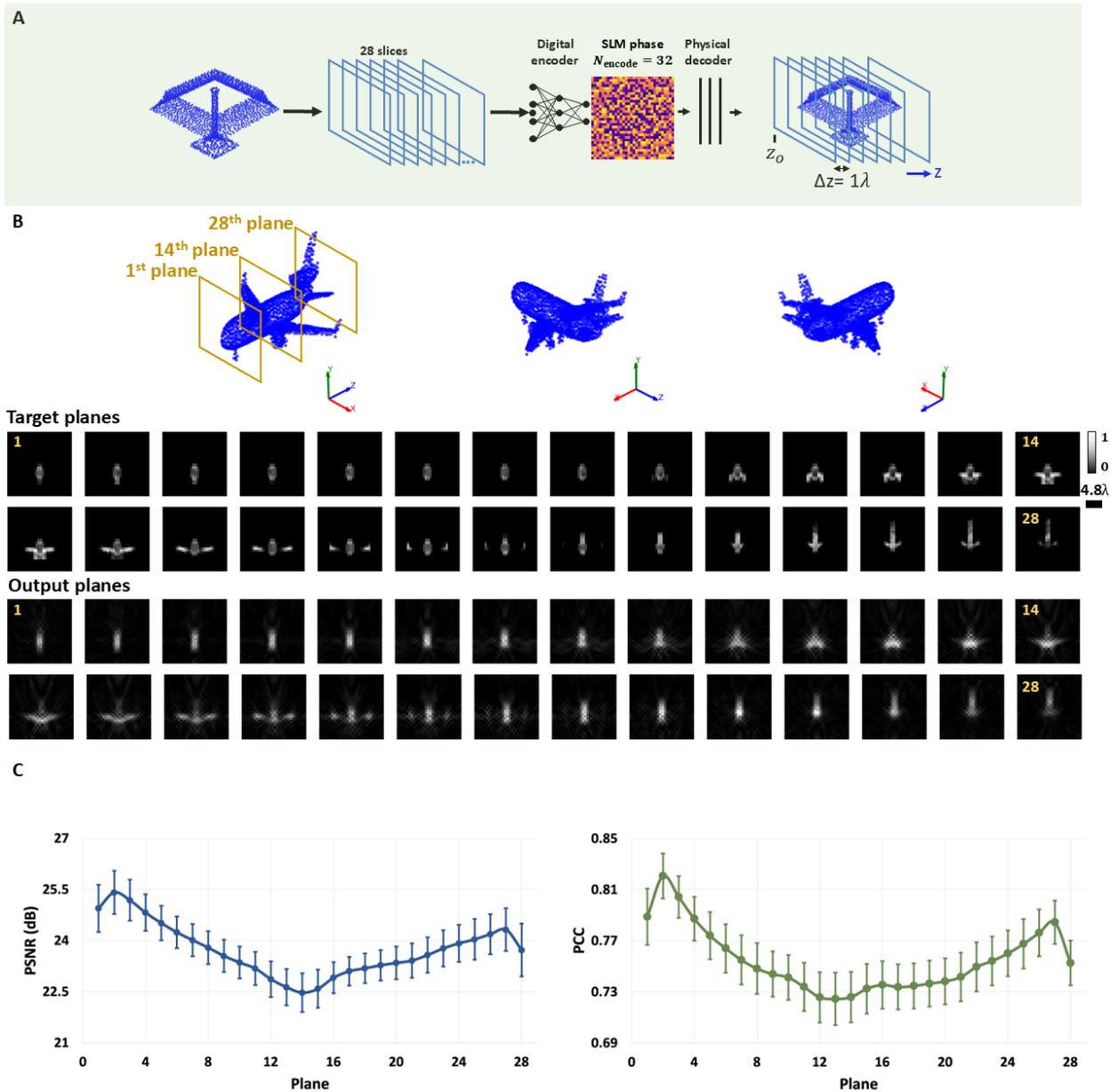

**Figure 5. Volumetric multi-slice projection using the snapshot diffractive 3D display. A**, Schematic of the 28-slice volumetric input and projection process using a digital encoder and a diffractive decoder, where a stack of axial slices was encoded into a single phase pattern and projected in a snapshot with a uniform axial spacing of $\Delta z = 1\lambda$. **B**, Target slices and corresponding projected outputs at selected axial positions. **C**, PSNR and PCC values were evaluated across all 28 axial planes, each with an axial separation of $\Delta z = 1\lambda$.

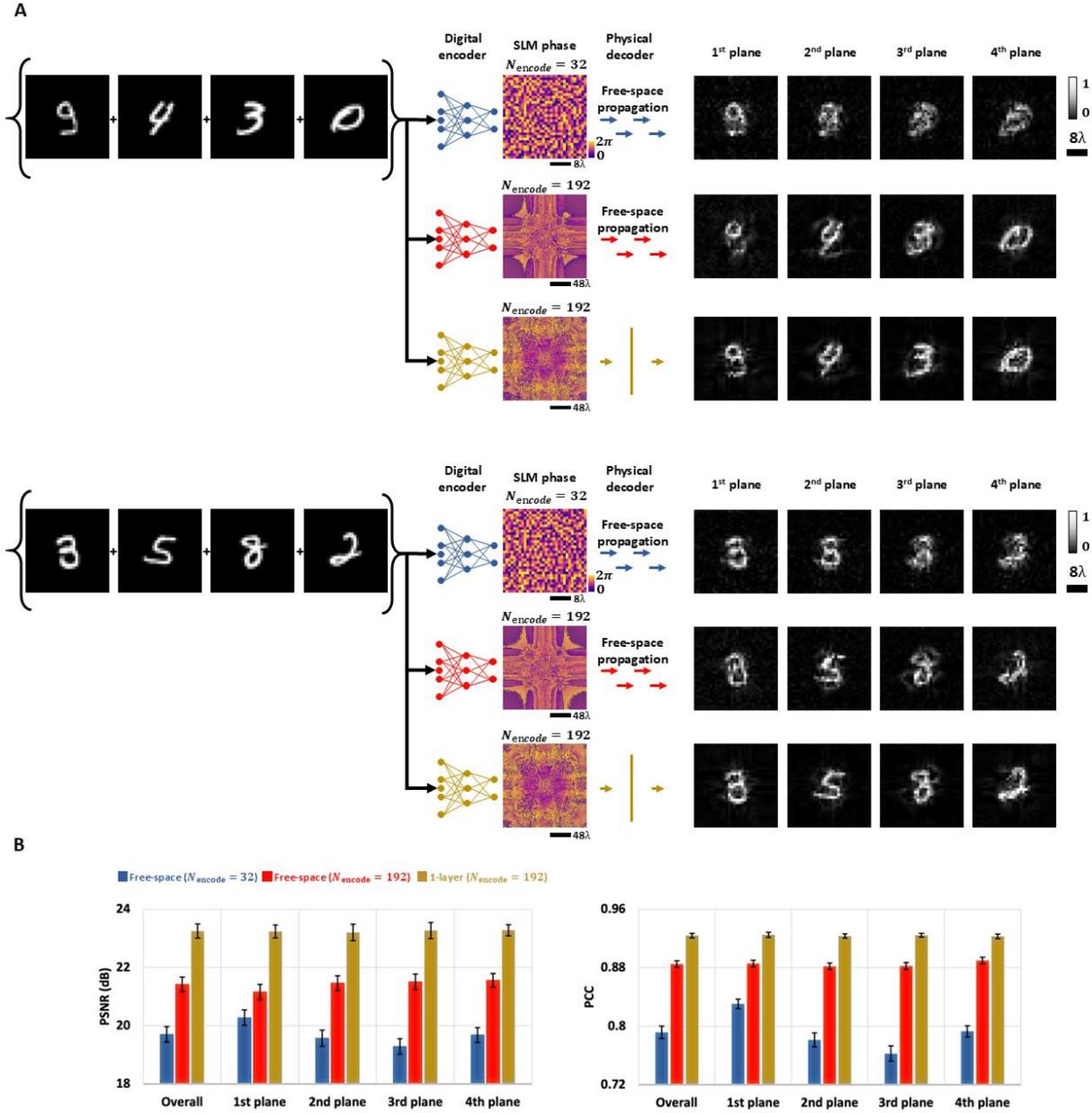

**Figure 6. Impact of the number of pixels used for the encoder phase pattern. A**, Representative multi-plane snapshot image projection examples obtained using two sampling resolutions, denoted as $N_{encode} \times N_{encode}$ with $N_{encode} \in \{32, 192\}$ (i.e., 32×32 and 192×192), evaluated under free-space propagation and a single-layer diffractive decoder. Increasing the number of encoder pixels increases the effective space–bandwidth product of the encoded wavefront, enabling a larger usable field of view, improved visual clarity, and reduced inter-plane cross-talk. **B**, PSNR and PCC values were evaluated across all axial planes as a function of $N_{encode}$, showing consistent fidelity improvements with increased pixel count and superior performance of the diffractive decoder compared to free-space propagation.

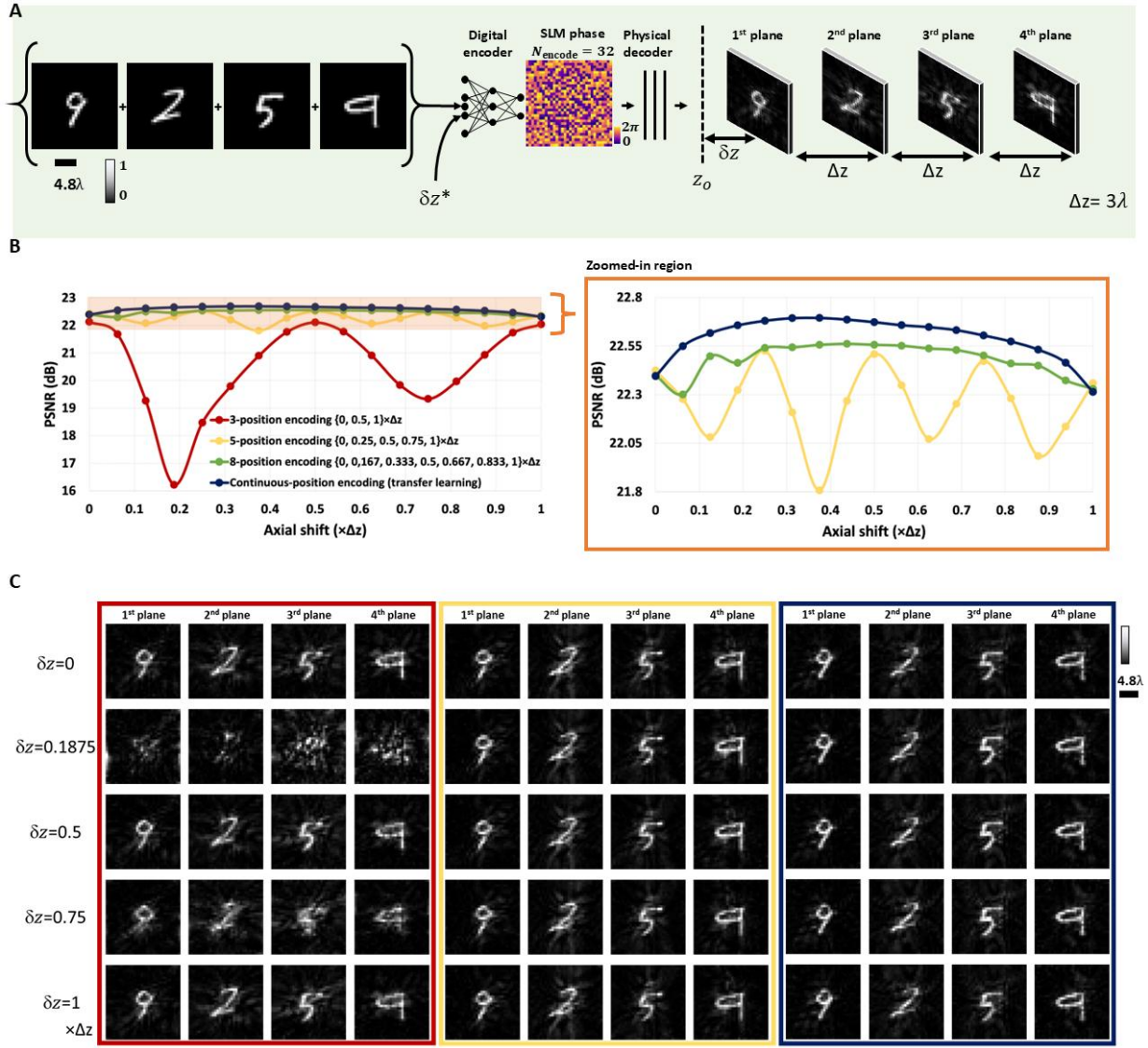

**Figure 7. Axial-shift adjustment ability of the snapshot diffractive 3D display under different axial encoding strategies. A**, Schematic of the distance adaptive diffractive 3D display, where an axial shift $\delta z^*$ was fed into the digital encoder to adjust the starting axial position of the projected display stack. **B**, Output PSNR as a function of the input-controlled axial shift $\delta z$ during testing for models trained with 3-position, 5-position, 8-position, and continuous-position encoding. Denser axial sampling during the training produced smoother PSNR curves as a function of image depth. **C**, Representative multi-plane snapshot image projections at different values of $\delta z$.

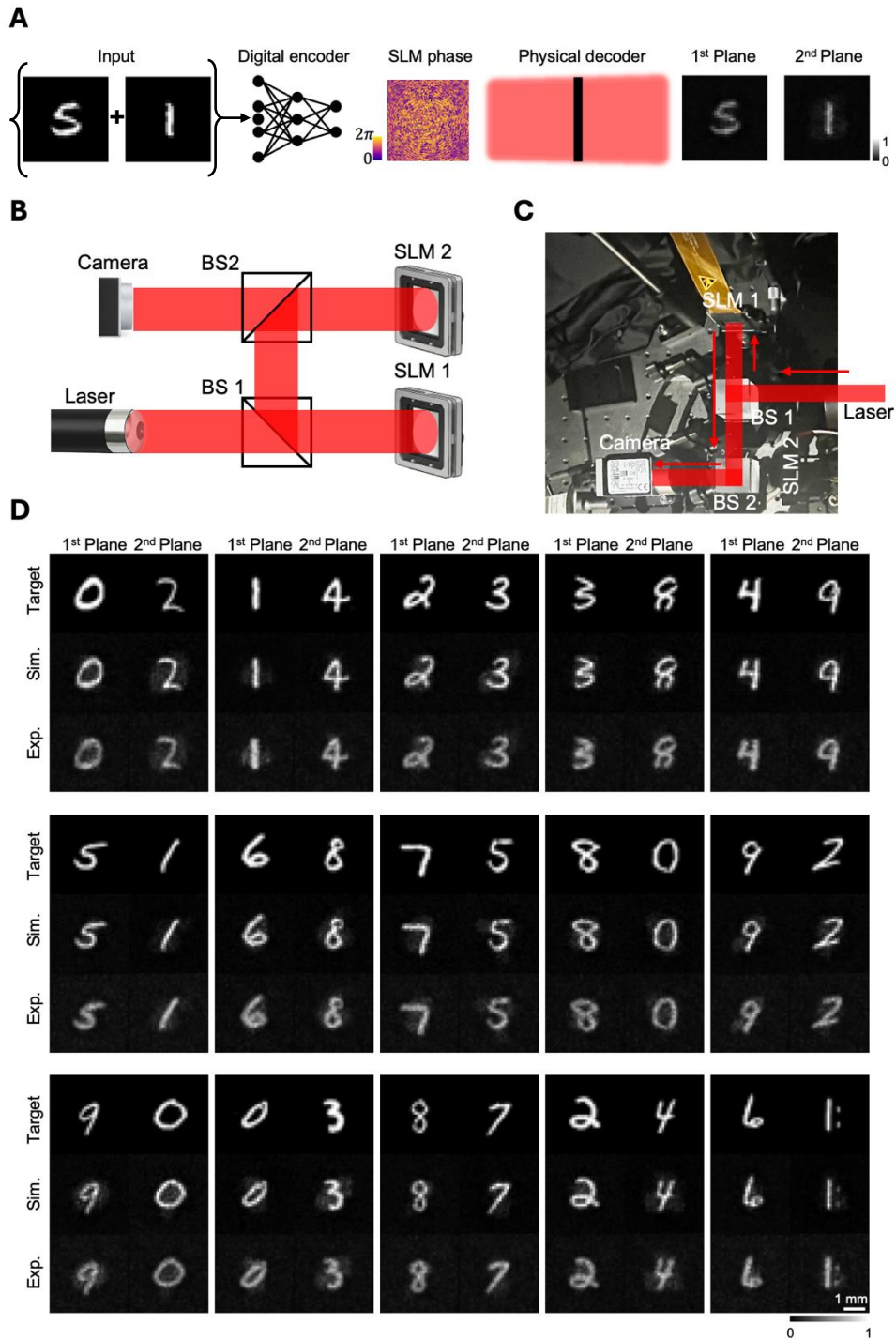

**Figure 8. Experimental validation of the diffractive snapshot 3D display system**. **A**, Schematic of the two-plane projection setup. **B**, Optical layout of the experimental system. **C**, Photograph of the physical setup showing the laser source, beam splitters, two reflective SLMs, and the imaging camera. **D**, Experimental results for multiple input test objects (never seen before). For each sample, target images (top), simulated numerical results (middle), and measured results (bottom)

are shown at the 1$^{st}$ and 2$^{nd}$ planes ($\Delta z = 1$ cm). BS: Beam Splitter, SLM: Spatial Light Modulator. Also see **Supplementary Fig. S1** for additional experimental results that further support the same conclusions.